# GRAVITATIONAL-WAVE DETECTOR "DULKYN"

*R. A. Daishev, S. M. Kozyrev, A. V. Lukin, A. I. Lubimov, S. V. Mavrin,
Z. G. Murzakanov, B.P. Pavlov, A. F. Skochilov, Yu. P. Chugunov*

The article deals with the results of the research on the realization of the scientific and technical project "Dulkyn" for detection of low frequency periodic gravitational radiation from the double relativistic astrophysical objects, done by the Scientific Centre of Gravitational-Wave Research "Dulkyn" of the Academy of Sciences of the Tatarstan Republic (SC GWR "Dulkyn" AS RT), and the Federal State Unitary Enterprise and R and D Corporation called "State Institute of Applied Optics" (FSUE NPO GIPO).

## Introduction

At the beginning of the 21$^{st}$ century the problem of gravitational phenomena made all the world leading specialists in the sphere of theoretical and experimental physics join their efforts. At the present time there are fourteen big research centres in the leading countries around the world which are conducting experiments aimed to detect gravitational radiation. They are: the USA (project LIGO), France-Italy (project VIGRO) Germany- Great Britain (project GEO-600), Japan (project TAMA-300), the international project of space basing called LISA, etc.

The experiments on detection of gravitational waves (GW) are considered to be an extremely difficult task. Practically the experiments of this kind have an objective to search for the most sensitive methods of detection, to define the sources of noise and interference and the way to reduce and compensate for them. On the strength of the anomalous weakness of GW signals measuring equipment with unique sensitiveness has to be selective about the GW information.

The development of the experimental base of gravitational-wave physics is directed at detection of gravitational radiation (GR) which must have a reliable evidence of its gravitational-wave nature, identification of the source of radiation and measurement of the characteristics of its astrophysical processes according to the structure of GW signal [1].

The large-scale receiving big-base optic-interferometer antennas that are being constructed now – the modified Michelson's interferometers – use mirrors in "free" masses as the elements which are sensitive to the influence of GW. The rated sensitiveness of the antennas to the stretch amplitude of GW is proportional to the effective interferometer arm length and amounts to $10^{-19} \div 10^{-20}$. These antennas are intended for detection of short impulse GW signals from flashing sources with unknown space-time characteristics, which decreases authenticity of their detection. The fact that there are a great number of models of GW signals from this kind of sources also makes it difficult to provide their optimal receiving. For a clear detection of GW signals from flashing sources the needed instant signal-noise relationship should exceed 1.

In Michelson's big-base interferometer the change in distance between the mirrors in the "free" masses in the field of GR is a useful effect. The outer interference (mechanic, acoustic, seismic, etc.), that influence the "free" masses, have the same effect. That is why the auto compensation for interference effects on the "free" masses (the absolute auto compensation for their oscillations) may also lead to simultaneous elimination of GW effects on them from the found source of GR. This circumstance makes it difficult to provide the needed signal-noise relationship.

According to the objectives of the experimental base of gravitational-wave physics, detection of GR of cosmic origin requires creation of widely-spread global net of surface GW detectors. Functioning of the GW antennas net is based on the scheme of coincidences with optimization of net orientation. Hard antennas of the Weber type and Michelson's big-base interferometers will allow us to make an optimal antenna array and organize the global GW service for flashing sources.

There are also periodic low-frequency GW signals from relative double astrophysical objects, whose space-frequency-time characteristics are known exactly. In spite of the fact that dimensionless amplitude of GR from these sources is two or three units less than the amplitude from flashing sources, and the GW antenna receives the detected GW signal mixed with GW background and GW signals from the other double astrophysical objects in the close frequency range, a priori information of the characteristics of the studied GW signals from the double systems, duration of their existence (up to $10^{13}$ s) will allow us to provide the optimal processing of GW signals coming from them, including their protracted (up to $10^{7}$ s) accumulation.

The International project of space basing LISA – a big-base (the arm length is 5 million km) interferometer – which is meant to detect low-frequency GR from double relative astrophysical objects, will be accomplished by 2015.

Since 1994 the analogous research has been conducted in Kazan within the framework of the scientific-technical project called "Dulkyn" (the Tatar word meaning 'wave') by the staff of United Experimental Laboratory of Gravitational-Optic Research (UELGOR) [2], which was made on the basis of State Institute of Applied Physics and Scientific Centre of Gravitational-Wave Research "Dulkyn" of the Academy of Sciences of the Tatarstan Republic (SC GWR "Dulkyn" AS RT). The project is worked out by SC GWR "Dulkyn" AS RT and intends to create a unique laser-interferometer complex to detect the periodic gravitational radiation from double relative astrophysical objects.

The GW detector "Dulkyn" is an antenna made of independent compact laser systems (GW antennas), constructed within the frame of one conception but presented in different variants (with triangular and pentagonal resonator configurations, generating stationary and travelling waves with and without the Sanjak's effect). Every GW antenna is a double-resonator laser system (DLS) with a nonequivalent resonators' geometry, common mirrors inflexibly fixed on one basis, and hologram diffraction arrays. The GW detector "Dulkyn", as well as the space basing detector in the LISA project, is intended for detection of low-frequency periodic signals. The GW detector "Dulkyn" is based on the other conception of formation of the effective signal. This conception uses the effect of immediate influence of GR on the optic radiation phase, which is modulated by the GW signal, and this effect prevails over the phase modulation, conditioned by GW evolution of the system of elastically fixed mirrors. GW influence on inflexibly fixed mirrors is weakened at least by a gear ratio $v/c \ll 1$ (v is velocity of sound in the environment, $c$ is velocity of light) [3].

The DLS of pentagonal and triangular configurations of the GW detector "Dulkyn" should be based on a hard fundament with no more than 2 meters in diameter. Fixing them on one pendant inside a vacuum cell will provide them with equal conditions for outer interference influence. In this case the coincidence of the signals at the exits of both laser systems of GW antennas, constructed within the frame of one conception but presented in different variants, will let us solve the problem of detection of GR and find the evidence of its GW nature and identify the source of its radiation provided that the GW signal is accumulated during a year.

**1. DLS of Triangular Configuration**

Elaboration of the conception of laser-interferometer detection of low-frequency GR within the framework of the project "Dulkyn" resulted in the following fundamental principles [4-6]:
– the detector of GR must be compact and placed on one basis;
– the interferometer must be active;
– the interferometer must have two geometrically nonequivalent resonators, which generate optic radiation with mutually orthogonal linear polarizations;

- the basic optical elements, including the active environment, must belong to both resonators, so that phase noise in the first and second measuring canals, caused by the elements, is correlated.

Owing to the usage of hologram diffraction reflecting elements all the principles have been implemented in the GR detectors, based on the DLS of pentagonal [4-5] and triangular configurations [7]. It has proved the vitality of the "Dulkyn" project conception.

Double pulsars emanate GR in the frequency range from $10^{-5}$ to $10^{-3}$ hertz. Periodic signals can occur in the same frequencies in the ring laser systems as a result of the Sanjak's effect; they are caused by irregularity of the Earth rotation, and constitute another problem for research. In order to clearly separate the signal of GR from the signal, conditioned by the Sanjak's effect, it is necessary to use a linear double-resonator laser with stationary waves, where the Sanjak's effect is absent, together with the ring pentagonal double-resonator laser[4-5] with travelling waves. The study [7] concerns the principles of the work and leveling of generation of optical radiation on stationary waves in the double-resonator linear laser, and is illustrated by means of the DLS of triangular configuration.

Fig. 1 demonstrates a variant of a linear laser with two resonators nonequivalent in terms of space which are constructed on the same optical elements. Both resonators have common active environment (AE) and are made of three hologram diffraction reflecting arrays 1, 2, 3 (volumetric phase transmitting arrays drawn on the mirror surface) located at the vertices of the equilateral triangle. Space frequencies in diffraction arrays are selected according to automatic collimation of diffraction of the minus first order (diffracted radiation is opposed in direction to the falling one). The method of computation of reflection factors of such hologram selectors for TE and TM polarizations of the light falling on them is considered in [8]. With the certain depth of modulation of dielectric penetration inside a volumetric phase array 2, TE radiation of polarization will diffract in the minus first order of diffraction, and TM radiation of polarization – in the zeroth order. In case of the array 1 we should satisfy the opposite condition. The array 1 will function as a mirror for a linear resonator 1 (1-2-3), and the array 2 – for the resonator 2 (2-1-3). The array 3 must have the maximum factor of reflection in the minus first order for TE and TM polarizations. Thus, a stationary electromagnetic wave of TE polarization will be generated in in the resonator 1, and a stationary wave of TM polarization – in the resonator 2.

As is shown in [8], hologram reflecting elements can have reflection factors up to 0.97 for every polarization in a proper diffraction order. A small percentage of radiation of the opposed polarization, penetrating in the 'wrong' circuit, is removed from resonators by means of $P_{TE}$ and $P_{TM}$ polarizers which constitute polarization prisms of the Glan's type and transmit linearly polarized light only of TE and TM polarizations correspondingly. The diaphragms D1 and D2 emit transverse $TEM_{00}$ modes in every resonator. Besides, we can greatly modify the degree of space overlapping in the active environment of the generated modes of TE and TM polarizations by means of displacement of one of the diaphragms perpendicularly to the primary position of the optical axis of the corresponding resonator; which leads to the sharp decrease in competition and relationship between them (this effect was first verified experimentally in [9]).

Fig. 1 shows that all the hologram diffraction reflective arrays 1, 2, 3 and the active environment belong to both resonators at the same time, as in the case of pentagonal ring double-resonator laser [4-5].

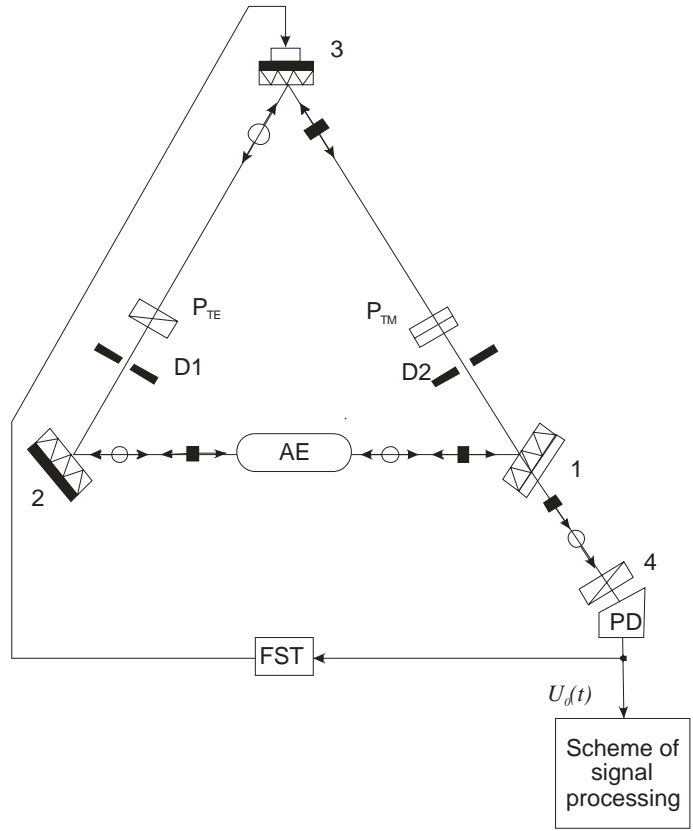

**Fig. 1. DLS of Triangular Configuration**

Generation equations and the method of computation of natural transverse frequencies of resonators in GR field have been considered in [5-7]. Replacement of natural transverse frequencies of the resonators 1 and 2 for the DLS of triangular configuration, shown in fig. 1, is caused by upheaval of gravitational wave perpendicularly to interferometer plane, and equals to:

$$\Delta\Omega_{1,2}(t) = \frac{3}{4}\Omega_{1,2}^0 \left(h_+ \pm \frac{\sqrt{3}}{3}h_\times\right)\sin(\Omega_g t + \beta_g), \qquad (1)$$

where $\Omega_{1,2}^0$ are natural frequencies of the resonators 1 и 2 without GR; $h_{+,\times}$ are stretch amplitudes of the first and second GW polarizations with TT calibration (the corresponding natural direction of the first polarization is meant to coincide with the direction of section 2-1 to simplify the computations); $\Omega_g$ и $\beta_g$ - are frequency and primary GW phase.

Within the synchronization zone ($\Omega_{1,2}^0 = \omega_0$) the variable part of phase difference $\Delta\varphi(t)$ of optical radiations phases of the resonators 1 and 2 looks like:

$$\Delta\varphi(t) = s(t) + n(t), \qquad (2)$$

$$s(t) = \Phi_g \sin(\Omega_g t + \beta_s), \quad \Phi_g = \frac{h_0 \omega_0}{\sqrt{\Omega_0^2 + \Omega_g^2}},$$

$$\beta_s = \beta_g - \arctan\left(\frac{\Omega_g}{\Omega_0}\right), \quad h_0 = \frac{\sqrt{3}}{2}h_\times,$$

where *s(t)* is a useful signal, *n(t)* is phase noise, caused by technical and natural fluctuations of different frequencies, $\Omega_0$ is coverage zone width.

Since $|s(t)| << |n(t)|$, there arises the task to minimize noise *n(t)* which can be done in a certain frequency range with the help of the system of stabilization for signal destruction (SS) [5,6,10].

The stabilization system of difference of optical radiation phases generated in two DLS resonators consists of an electronic block of frequency-phase self-tuning (FST) (fig. 1). Nonworking diffraction orders of hologram diffraction array 1 are used to let out optical radiation from DLS resonators 1 and 2. Overlapped by means of rays superposition scheme orthogonal irradiations can create interference field registered by photodetector (PD) after passing through the polarizer 4, whose plane for transmission of light with linear polarization makes an angle of 45 ° with picture plane. An error signal that controls the work of piezoelement fixed on the mirror 3 is formed according to the output voltage of PD in FST. Piezoelement provides maintenance of the regime of radiation synchronization in both inner and outer resonators. Here it forms the error signal controlling the work of a special phase modulator [5, 6], that maintains phase difference within the limits of the given level of threshold of stabilization.

As a result of work of SS the spectral density of amplitude of the reduced phase noise $S_n(f)$ will not exceed the level $S_n^0 = 10^{-4} \, rad/\sqrt{Hz}$ in the frequency range 0<*f*<10 (Hz). Under these conditions we evaluate the time necessary to accumulate T in order to detect low frequency periodic GI with specific parameters $h_0=10^{-22}$ и $\Omega_g/2\pi=10^{-3}$ Hz. Provided that $\Phi_g \geq qS_n(f)/\sqrt{T}$, where $q \approx 3$ is signal / noise relationship necessary to detect GW signal, (2) leads us to $\sqrt{T} \approx S_n^0 \sqrt{\Omega_0^2 + \Omega_g^2}/\omega_0 h_0$, whence, in case the frequency of optical irradiation equals $\omega_0/2\pi \approx 10^{15}$ Hz, and the coverage zone width equals $\Omega_0/2\pi \approx 1$ Hz, the time is T≈$10^7$c≈4 months. It should be noticed that so long as the coverage zone width is much bigger than frequency of any GR of the type PSR J1537+1155, the time needed for the signal detection depends upon the dimensionless amplitude of GW signal only, and doesn't depend on its period.

Unfortunately, detection of the sought for GW signal based on the examination of noises spectrum doesn't define the form and phase of the periodic signal. The information of the phase upheaval of the detected GW signal, which happens when GW detector, together with the Earth, orbits around the sun, can bring us to a separate definition of direction to the source of GR [11].

Definition of form and computation of the phase upheaval of the received periodic signal in the GW detector "Dulkyn" are stipulated by systems of inter periodic and inner periodic accumulation of the actual signal. The main principle of their functioning is described in detail in the works [6, 12-14]. There was offered a method of separation of infra low frequency signal from noises in order to reduce the time necessary for the certain detection of GW signal [15].

### Experimental results

The main experimental base of UELGOR is the specialized premises of FSUE NPO GIPO provided with all the necessary equipment.

The experiments are conducted in a specialized cabin, placed in a subsurface laboratory of diffraction arrays (SLDA) on the depth of 12 meters from the earth surface in order to minimize influence of vibration phone and simplify the task of temperature stabilization. Walls and floor of the laboratory are hydroisolated and placed in an all-metal capacity which rests upon a sand cushion. The laboratory has a technical cellar that is 2 meters deep and is used to set up the basement of the main equipment of the SLDA and place electric communications and systems providing the basic equipment with electricity. The experimental cabin is made of two isolated cabins-cells. All the high-sensitiveness equipment in the inner cabin, including the workable detector, is set up on a special basement, which is, in its turn, isolated from vibrations coming from the foundation of the premises. The fundament is a ferroconcrete block weighing 100 tons which is vibroisolated by means of spiral wagon springs. A supplementary vibroisolation of the equipment is implemented by special vibroisolators.

A three-stage system of thermostabilization creates the necessary temperature conditions. At the first stage an air-conditioner heats the stream of air coming to SLDA as a supply air conditioner and thus maintains the stable temperature within the range of 0.5 °C (0.25 °C from the given indicated value). The second and the third stages are provided by a specially developed autonomous system of temperature stabilization. The temperature stabilization in the intercabin space within the range of tenths of a degree of the indicated value takes place at the second stage; and, correspondingly, the temperature stabilization in the inner cabin within the range of hundredths of a degree takes place at the second.

The first step to the implementation of the complex programme of gravitational experiments carried out by the Project "Dulkyn" was to make a passive variant of GW detector based on the pentagonal ring double-circuit interferometer [16,17], with the source of optical radiation outside the interferometer.

In December, 1995, a pentagonal interferometer, working at the wave length of 0.63 micrometers, was assembled and adjusted on a round polished glass plate 10sm thick, 65 sm in diameter, and one side of the pentagon was 30 sm long. The following technical tasks have been accomplished in the course of the test experiments with the passive interferometer.

1. We have elaborated and implemented the original system of methods for adjustment of a complex double-circuit pentagonal optical resonator, constructed on the common optical elements, two of which are hologram diffraction arrays, made specially in UELGOR. The experiments conducted under the conditions of strong mechanical interference showed that pentagon, as a double-circuit optical configuration, is steady and doesn't need a subsidiary adjustment monitoring. We have designated the degree of identity of the first and the second hologram reflection elements. The present methods have been wholly used in construction of an active variant of GW detector, i. e. the pentagonal DLS.

2. We have tested different schemes of formation of interference field at the detector outlet, and the corresponding methods for processing of the interference picture; and maximally used the double-circuit of the interferometer and the opportunity to create two information canals.

3. We have modeled and tested a system of correlation autocompensation for interferences (SCAI) under the conditions of real interferences and noise. We have determined the threshold of sensitiveness of the SCAI for the passive interferometer. We have come to the conclusion that the SCAI of this type can be used for the primary rough filtration of strong correlated interferences.

4. With the help of the passive interferometer we have created and worked up the original signal breaking SS for phase differences of optical radiations in the first and second circuits of the pentagon, which allowed us to reduce phase noise in the low frequency area by several orders.

5. We have researched for the real spectral density of phase noise in the infra low frequencies (a stochastic component of phase difference in the first and second canals of the interferometer) in the range of $10^{-5} \div 10^{-1}$ Hz (the working range for detectors of periodic gravitational radiation from double relative astrophysical objects). For this purpose we have carried out a series of continuous long-run measurements with stabilization system being turned off and on. Thereafter, we have received spectral contours for both reduced and unreduced noise.

6. We have conducted a series of experiments with analog and digital simulation of GW influence. Periodic signal with the amplitude of $10^{-4} \div 10^{-6}$ rad was transmitted to the signal circuit, mixed with real interference signals (in working and nonworking stabilization system), and then separated from the outlet signal by means of specially developed threshold algorithm. We have found the ratio between signal amplitude, stabilization threshold and level of unreduced noise, which provide the "criterion for recognition" of simulated periodic signal, for the passive interferometer.

7. We have elaborated and tested methods of inter periodic and inner periodic accumulation of periodic signals, with the help of which we can confidently distinguish a weak low frequency useful signal on the background of strong interferences.

8. We have fulfilled a complex checkup of operational capability of the whole working "simulation – recognition" cycle of periodic signal in the infra low frequency in the automated regime with cycles of 2 to 8 days and nights. As a result we could test the system of computer control over the detection regime, which will be used when the active variant of the GW detector is in operation.

The second step to the implementation of the Project "Dulkyn" is to create a prototype of the active variant of the GW detector, i.e. a compact double-resonator laser system.

We have assembled and adjusted the DLS of pentagonal ring configuration and triangular linear configuration in the summer, 1999. Using these schemes we received steady generation on the wavelength of 3.39 micrometers.

All the optical elements making up the experimental double-resonator laser system were fixed inflexibly on one basement. For this purpose we used the foundation-plane of 1.4 m in diameter and 8 sm thick, made of "E" – alloy D-16 (with mass of 400 kg), undergone a special thermal processing to stabilize its dimensions. Thermal processing included four regimes of heating and cooling in different temperatures to remove inner voltage from the metal.

We have designed a specialized vacuum cell (VC) of 1.7 m in diameter and 0.67 m high (with mass of 550 kg), which is meant for isolation of the DLS from the environment. The residual pressure inside the vacuum cell does not exceed $10^{-3}$ Tor.

The foundation-plane of the DLS lies freely inside the VC on three bearings, placed at the vertexes of the regular triangle and apart from the centre of the plane by 2 thirds of its radius.

Nowadays we carry out the work on adjustment and setup of service systems of frequency and phase stabilization of the generated optical radiation in the prototypes of the active variant of the GW detector.

## Conclusion

Preliminary experimental research, done by UELGOR, showed that there is an opportunity to create an optical part of the GW detector "Dulkyn" basing on the compact double-resonator laser system. We are planning to do an experimental research into calibration of the GW detector on completion of the checkup of operational capability of the DLS in vacuum conditions.